\documentclass[10pt,twocolumn,letterpaper]{article}

\usepackage{cvpr}
\usepackage{times}
\usepackage{epsfig}
\usepackage{graphicx}
\usepackage{amsmath}
\usepackage{amssymb}
\usepackage{url}

\usepackage{hyperref}

\cvprfinalcopy 

\def\cvprPaperID{****} 

\begin{document}

\title{A Vision-based Social Distancing and Critical Density Detection System for COVID-19}

\author{
Dongfang Yang\thanks{Equal contribution} \hspace{1.8cm} 
Ekim Yurtsever\footnotemark[1] \hspace{1.8cm}
Vishnu Renganathan\\
Keith A. Redmill\hspace{1.8cm} 
\"{U}mit \"{O}zg\"{u}ner \\
\\
The Ohio State University, Columbus, OH 43210, USA \\
\texttt{\small{\{yang.3455,yurtsever.2,renganathan.5,redmill.1,ozguner.1\}@osu.edu}}\\
}

\maketitle








%

\begin{abstract}
Social distancing has been proven as an effective measure against the spread of the infectious COronaVIrus Disease 2019 (COVID-19). However, individuals are not used to tracking the required 6-feet (2-meters) distance between themselves and their surroundings. An active surveillance system capable of detecting distances between individuals and warning them can slow down the spread of the deadly disease. Furthermore, measuring social density in a region of interest (ROI) and modulating inflow can decrease social distancing violation occurrence chance. 

On the other hand, recording data and labeling individuals who do not follow the measures will breach individuals' rights in free-societies. Here we propose an Artificial Intelligence (AI) based real-time social distancing detection and warning system considering four important ethical factors: (1) the system should never record/cache data, (2) the warnings should not target the individuals, (3) no human supervisor should be in the detection/warning loop, and (4) the code should be open-source and accessible to the public. Against this backdrop, we propose using a monocular camera and deep learning-based real-time object detectors to measure social distancing. If a violation is detected, a non-intrusive audio-visual warning signal is emitted without targeting the individual who breached the social distancing measure. Also, if the social density is over a critical value, the system sends a control signal to modulate inflow into the ROI. We tested the proposed method across real-world datasets to measure its generality and performance. The proposed method is ready for deployment, and our code is open-sourced~\footnote{\url{https://github.com/dongfang-steven-yang/social-distancing-monitoring}}.   
\end{abstract}

\section{Introduction}


Social distancing is an effective measure~\cite{courtemanche2020strong} against the novel COronaVIrus Disease 2019 (COVID-19) pandemic. However, the general public is not used to keep an imaginary safety bubble around themselves. An automatic warning system~\cite{nguyen2020enabling, agarwal2020unleashing, punn2020monitoring, cristani2020visual} can help and augment the perceptive capabilities of individuals. 

Deploying such an active surveillance system requires serious ethical considerations and smart system design. The first challenge is privacy~\cite{chan2008privacy, qureshi2009object, park2005track}. If data is recorded and stored, the privacy of individuals may be violated intentionally or unintentionally. As such, the system must be real-time without any data storing capabilities.

Second, the detector must not discriminate. The safest way to achieve this is by building an AI-based detection system. Removing the human out of the detection loop may not be enough -- the detector must also be design free. Domain-specific systems with hand-crafted feature extractors may lead to malign designs. A connectionist machine learning system, such as a deep neural network without any feature-based input space, is much fairer in this sense, with one caveat: the distribution of the training data must be fair. 


Another critical aspect is being non-intrusive. Individuals should not be targeted directly by the warning system. A non-alarming audio-visual cue can be sent to the vicinity of the social distancing breach to this end. 

\begin{figure*}[t]
    \begin{center}
        \includegraphics[width=1\linewidth]{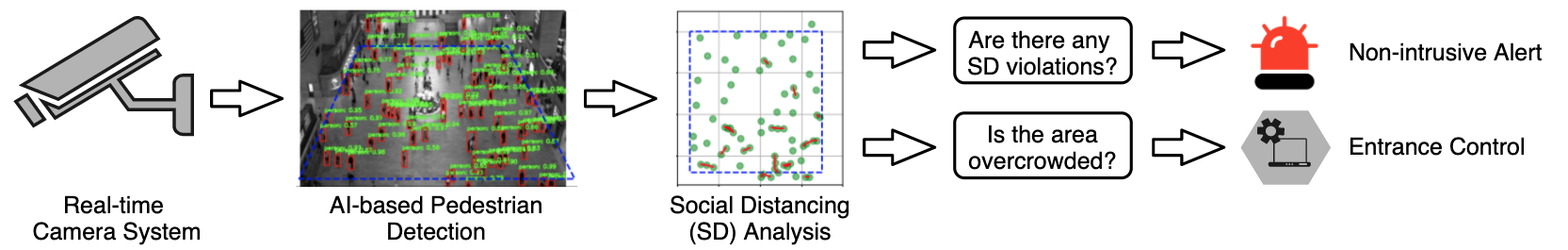}
    \end{center}
    \caption{Overview of the proposed system. Our system is real-time and does not record data. An audio-visual cue is emitted each time an individual breach of social distancing is detected. We also make a novel contribution by defining a critical social density value $\rho_c$ for measuring overcrowding. Entrance into the region-of-interest can be modulated with this value. }
    \label{fig:overview}
\end{figure*}

The system must be open-sourced. This is crucial for establishing trust between the active surveillance system and society.

Against this backdrop, we propose a non-intrusive augmentative AI-based active surveillance system for sending omnidirectional visual/audio cues when a social distancing breach is detected. The proposed system uses a pre-trained deep convolutional neural network (CNN)~\cite{ren2015faster, bochkovskiy2020yolov4} to detect individuals with bounding boxes in a given monocular camera frame. Then, detections in the image domain are transformed into real-world bird's-eye view coordinates. If a distance smaller than the threshold is detected, the system emits a non-alarming audio-visual cue. Simultaneously, the system measures social density. If the social density is over a critical threshold, the system sends an advisory inflow modulation signal to prevent overcrowding. 
v

Our main contributions are:

\begin{itemize}
    \item A novel vision-based real-time social distancing and critical social density detection system 
    \item Definition of critical social density and a statistical approach to measuring it
    \item Measurements of social distancing and critical density statistics of common crowded places such as the  New York Central Station, an indoor mall, and a busy town center in Oxford.
\end{itemize}


\section{Related Work}


\textbf{Social distancing for COVID-19.} COVID-19 has caused severe acute respiratory syndromes around the world since December 2019~\cite{hui2020continuing}. Recent work showed that social distancing is an effective measure to slow down the spread of COVID-19 \cite{courtemanche2020strong}.  
Social distancing is defined as keeping a minimum of 2 meters (6 feet) apart from each individual to avoid possible contact. Further analysis~\cite{greenstone2020does} also suggests that social distancing has substantial economic benefits. COVID-19 may not be completely eliminated in the short term, but an automated system that can help monitoring and analyzing social distancing measures can greatly benefit our society.




\textbf{Pedestrian detection.} 
Pedestrian detection can be regarded as either a part of a general object detection problem or as a specific task of detecting pedestrians only. A detailed survey of 2D object detectors, as well as datasets, metrics, and fundamentals, can be found in~\cite{zou2019object}. Another survey~\cite{zhao2019object} focuses on deep learning approaches for both generic object detection and pedestrian detection. State-of-the-art object detectors use deep learning approaches, which are usually divided into two categories. The first one is called two-stage detectors, mostly based on R-CNN~\cite{girshick2014rich, girshick2015fast, ren2015faster, dai2016r}, which starts with region proposals and then performs the classification and bounding box regression. The second one is called one-stage detectors, of which the famous models are YOLOv1-v4~\cite{redmon2016you, redmon2017yolo9000, redmon2018yolov3, bochkovskiy2020yolov4}, SSD~\cite{liu2016ssd}, RetinaNet~\cite{lin2017focal}, and EfficientDet~\cite{tan2020efficientdet}. In addition to these anchor-based approaches, there are also some anchor-free detectors: CornerNet~\cite{law2018cornernet}, CenterNet~\cite{duan2019centernet}, FCOS~\cite{tian2019fcos}, and RepPoints~\cite{yang2019reppoints}. These models were usually evaluated on datasets of Pascal VOC~\cite{everingham2010pascal, everingham2015pascal} and MS COCO~\cite{lin2014microsoft}. The accuracy and real-time performance of these approaches are good enough for deploying pre-trained models for social distancing detection.


\textbf{Social distancing monitoring.}  
Emerging technologies can assist in the practice of social distancing. A recent work~\cite{nguyen2020enabling} has identified how emerging technologies like wireless, networking, and artificial intelligence (AI) can enable or even enforce social distancing. The work discussed possible basic concepts, measurements, models, and practical scenarios for social distancing. Another work~\cite{agarwal2020unleashing} has classified various emerging techniques as either human-centric or smart-space categories, along with the SWOT analysis of the discussed techniques. A specific social distancing monitoring approach~\cite{punn2020monitoring} that utilizes YOLOv3 and Deepsort was proposed to detect and track pedestrians followed by calculating a violation index for non-social-distancing behaviors. The approach is interesting but results do not contain any statistical analysis. Furthermore, there is no implementation or privacy-related discussion other than the violation index. Social distancing monitoring is also defined as a visual social distancing (VSD) problem in~\cite{cristani2020visual}. The work introduced a skeleton detection based approach for inter-personal distance measuring. It also discussed the effect of social context on people's social distancing and raised the concern of privacy. The discussions are inspirational but again it does not generate solid results for social distancing monitoring and leaves the question open. 

Very recently, several prototypes utilizing machine learning and sensing technologies have been developed to help social distancing monitoring. Landing AI~\cite{landingai2020} has proposed a social distancing detector using a surveillance camera to highlight people whose physical distance is below the recommended value. A similar system~\cite{khandelwal2020using} was deployed to monitor worker activity and send real-time voice alerts in a manufacturing plant. In addition to surveillance cameras, LiDAR based~\cite{hall2020social} and stereo camera based~\cite{zed2020using} systems were also proposed, which demonstrated that different types of sensors besides surveillance cameras can also help. 

The above systems are interesting, but recording data and sending intrusive alerts might be unacceptable by some people. On the contrary, we propose a non-intrusive warning system with softer omnidirectional audio-visual cues. In addition, our system evaluates critical social density and modulates inflow into a region-of-interest.


\section{Preliminaries}

\textbf{Object detection with deep learning.} Object detection in the image domain is a fundamental computer vision problem. The goal is to detect instances of semantic objects that belong to certain classes, e.g., humans, cars, buildings. Recently, object detection benchmarks have been dominated by deep convolutional neural networks (CNNs) models \cite{girshick2014rich, girshick2015fast, ren2015faster, dai2016r, redmon2016you, redmon2017yolo9000, redmon2018yolov3, bochkovskiy2020yolov4, liu2016ssd, lin2017focal}. For example, top scores on MS COCO~\cite{lin2014microsoft}, which has over 123K images and 896K objects in the training-validation set and 80K images in the testing set with 80 categories, have almost doubled thanks to the recent breakthrough in deep CNNs.

These models are usually trained by supervised learning, with techniques like data augmentation~\cite{zoph2019learning} to increase the variety of data. 


\textbf{Model Generalization.} The generalization capability~\cite{zhang2016understanding, liu2020understanding} of the state-of-the-art is good enough for deploying pre-trained models to new environments. For 2D object detection, even with different camera models, angles, and illumination conditions, pre-trained models can still achieve good performance. 

Therefore, a pre-trained state-of-the-art deep learning based pedestrian detector can be directly utilized for the task of social distancing monitoring.

\section{Method}

We propose to use a fixed monocular camera to detect individuals in a region of interest (ROI) and measure the inter-personal distances in real time without data recording. The proposed system sends a non-intrusive audio-visual cue to warn the crowd if any social distancing breach is detected. Furthermore, we define a novel critical social density metric and propose to advise not entering into the ROI if the density is higher than this value. The overview of our approach is given in Figure \ref{fig:overview}, and the formal description starts below.


\subsection{Problem formulation}

We define a scene at time $t$ as a 6-tuple $S = (\mathbf{I}, A_0, d_c, c_1,c_2, U_0)$, where $\mathbf{I} \in \mathbb{R}^{H\times W \times3}$ is an RGB image captured from a fixed monocular camera with height $H$ and width $W$. $A_0 \in \mathbb{R}$ is the area of the ROI on the ground plane in real world and $d_c \in \mathbb{R}$ is the required minimum physical distance. $c_1$ is a binary control signal for sending a non-intrusive audio-visual cue if any inter-pedestrian distance is less than $d_c$. $c_2$ is another binary control signal for controlling the entrance to the ROI to prevent overcrowding. Overcrowding is detected with our novel definition of critical social density $\rho_c$. $\rho_c$ ensures social distancing violation occurrence probability stays lower than $U_0$. $U_0$ is an empirically decided threshold such as 0.05.



\textbf{Problem 1.} Given $S$, we are interested in finding a list of pedestrian pose vectors $P = (\mathbf{p}_1, \mathbf{p}_2, \cdots , \mathbf{p}_n)$, $\mathbf{p} \in \mathbb{R}^2$, in real-world coordinates on the ground plane and a corresponding list of inter-pedestrian distances $D=( d_{1,2}, \cdots, d_{1,n}, d_{2,3}, \cdots, d_{2,n}, \cdots, d_{n-1,n})$, $d \in \mathbb{R}^+$. $n$ is the number of pedestrians in the ROI. Also, we are interested in finding a critical social density value $\rho_c$. $\rho_c$ should ensure the probability  $p(d>d_c|\rho < \rho_c)$ stays over $1-U_0$, where we define social density as $\rho := n/A_0$.




Once Problem 1 is solved, the following control algorithm can be used to warn/advise the population in the ROI.

\textbf{Algorithm 1.} If $d \leq d_c$, then a non-intrusive audio-visual cue is activated with setting the control signal $c_1=1$, otherwise $c_1=0$. In addition, If $\rho > \rho_c,$ then entering the area is not advised 
with setting $c_2=1$, otherwise $c_2=0$.   

Our solution to Problem 1 starts below.


\subsection{Pedestrian detection in the image domain}

First, pedestrians are detected in the image domain with a deep CNN model trained on a real-world dataset: 

\begin{equation}
 \label{eq:cnn}
\{T_i\}_k = f_{\text{cnn}}(\mathbf{I}).
\end{equation}

$f_\text{cnn}:\mathbf{I}\rightarrow \{T_i\}_n$ maps an image $\mathbf{I}$ into $n$ tuples $T_i=(l_i \mathbf{b}_i, s_i), \forall i \in \{1, 2, \cdots, n\}$. $n$ is the number of detected objects. $l_i \in L$ is the object class label, where $L$, the set of object labels, is defined in $f_\text{cnn}$. $\mathbf{b}_i = (\mathbf{b}_{i,1},\mathbf{b}_{i,2},\mathbf{b}_{i,3},\mathbf{b}_{i,4})$ is the associated bounding box (BB) with four corners. $\mathbf{b}_{i,j}= (x_{i,j}, y_{i,j})$ gives pixel indices in the image domain. The second sub-index $j$ indicates the corners at top-left, top-right, bottom-left, and bottom-right respectively. $s_i$ is the corresponding detection score. Implementation details of $f_\text{cnn}$ is given in Section \ref{sec:implementation}.

We are only interested in the case of $l=$ `person'. We define $\mathbf{p}'_i$, the pixel pose vector of person $i$, with using the middle point of the bottom edge of the BB:
\begin{equation}
 \label{eq:2dpose}
\mathbf{p}'_i := \frac{(\mathbf{b}_{i,3} + \mathbf{b}_{i,4})}{2}.
\end{equation}

\subsection{Image to real-world mapping}

The next step is obtaining the second mapping function $h: \mathbf{p}' \rightarrow \mathbf{p}$. $h$ is an inverse perspective transformation function that maps $\mathbf{p}'$ in image coordinates to  $\mathbf{p}\in\mathbb{R}^2$ in real-world coordinates. $\mathbf{p}$ is in 2D bird's-eye-view (BEV) coordinates by assuming the ground plane $z=0$. We use the following well-known inverse homography transformation~\cite{forsyth2002computer} for this task:
\begin{equation}
    \mathbf{p}^{\text{bev}}=\mathbf{M}^{-1}\mathbf{p}^{\text{im}},
    \label{eq:im2bev}
\end{equation}
where $\mathbf{M}\in \mathbb{R}^{3\times 3}$ is a transformation matrix describing the rotation and translation from world coordinates to image coordinates. $\mathbf{p}^\text{im}=[p_x', p_y', 1]$ is the homogeneous representation of $\mathbf{p}'=[p_x', p_y']$ in image coordinates, and $\mathbf{p}^\text{bev}=[p^\text{bev}_x, p^\text{bev}_y, 1]$ is the homogeneous representation of the mapped pose vector.


The world pose vector $\mathbf{p}$ is derived from $\mathbf{p}^\text{bev}$ with $\mathbf{p} = [p^\text{bev}_x, p^\text{bev}_y]$.

\subsection{Social distancing detection}

After getting $P = (\mathbf{p}_1, \mathbf{p}_2, \cdots , \mathbf{p}_n)$ in real-world coordinates, obtaining the corresponding list of inter-pedestrian distances $D$ is straightforward. The distance $d_{i,j}$ for pedestrians $i$ and $j$ is obtained by taking the Euclidean distance between their pose vectors:

\begin{equation}
    \label{eq:distance}
    d_{i,j}=\lVert\mathbf{p}_i-\mathbf{p}_j\rVert.
\end{equation}



And the total number of social distancing violations $v$ in a scene can be calculated by: \begin{equation}
    v= \sum_{i=1}^n \sum_{\substack{j=1\\j \neq i}}^n \mathbb{I}(d_{i,j}),
\end{equation}
where $\mathbb{I}(d_{i,j})=1$ if $d_{i,j} < d_c$, otherwise $0$.

\subsection{Critical social density estimation}

\begin{figure}
    \centering
    \includegraphics[width=0.79\linewidth]{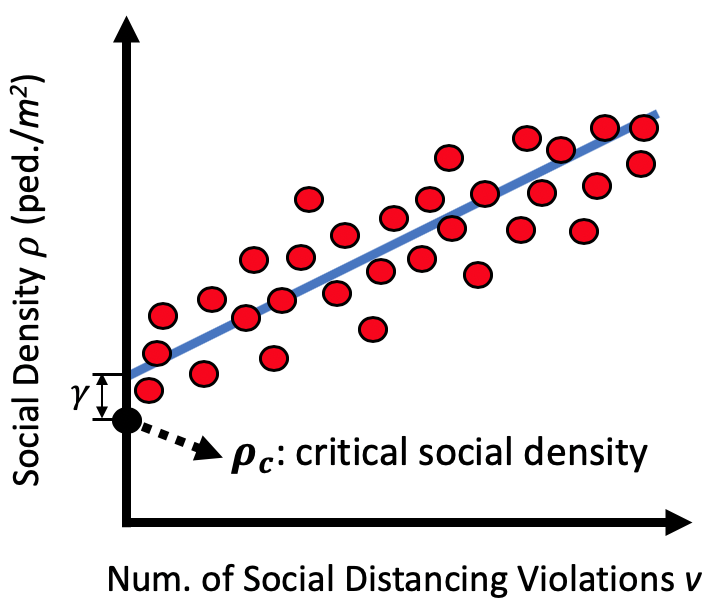}
    \caption{Obtaining critical social density $\rho_c$. Keeping $\rho$ under $\rho_c$ will drive the number of social distancing violations $v$ towards zero with the linear regression assumption.}
    \label{fig:y_intercept}
\end{figure}

Finally, we want to find a critical social density value $\rho_c$ that can ensure the social distancing violation occurrence probability stays below $U_0$. It should be noted that a trivial solution of $\rho_c = 0$ will ensure $v=0$, but it has no practical use. Instead, we want to find the maximum critical social density $\rho_c$ that can still be considered safe.  

To find $\rho_c$, we propose to conduct a simple linear regression using social density $\rho$ as the independent variable and the total number of violations $v$ as the dependent variable:

\begin{equation}
    v=\beta_0 + \beta_1\rho + \epsilon,
    \label{eq:regression}
\end{equation}
where $\boldsymbol{\beta}=[\beta_0,\beta_1]$ is the regression parameter vector and $\epsilon$ is the error term which is assumed to be normal. The regression model is fitted with the ordinary least squares method. Fitting this model requires training data. However, once the model is learned, data is not required anymore. After deployment, the surveillance system operates without recording data. 

Once the model is fitted, critical social density is identified as:
\begin{equation}
    \rho_c=\rho_\text{lb}^\text{pred},
\end{equation}
where $\rho_\text{lb}^\text{pred}$ is the lower bound of the $95\%$ prediction interval $(\rho_\text{lb}^\text{pred}, \rho_\text{ub}^\text{pred})$ at $v=0$, as illustrated in Figure~\ref{fig:y_intercept}.

Keeping $\rho$ under $\rho_c$ will keep the probability of social distancing violation occurrence near zero with the linear regression assumption.

\begin{figure*}
    \centering
	\includegraphics[width=0.85\linewidth]{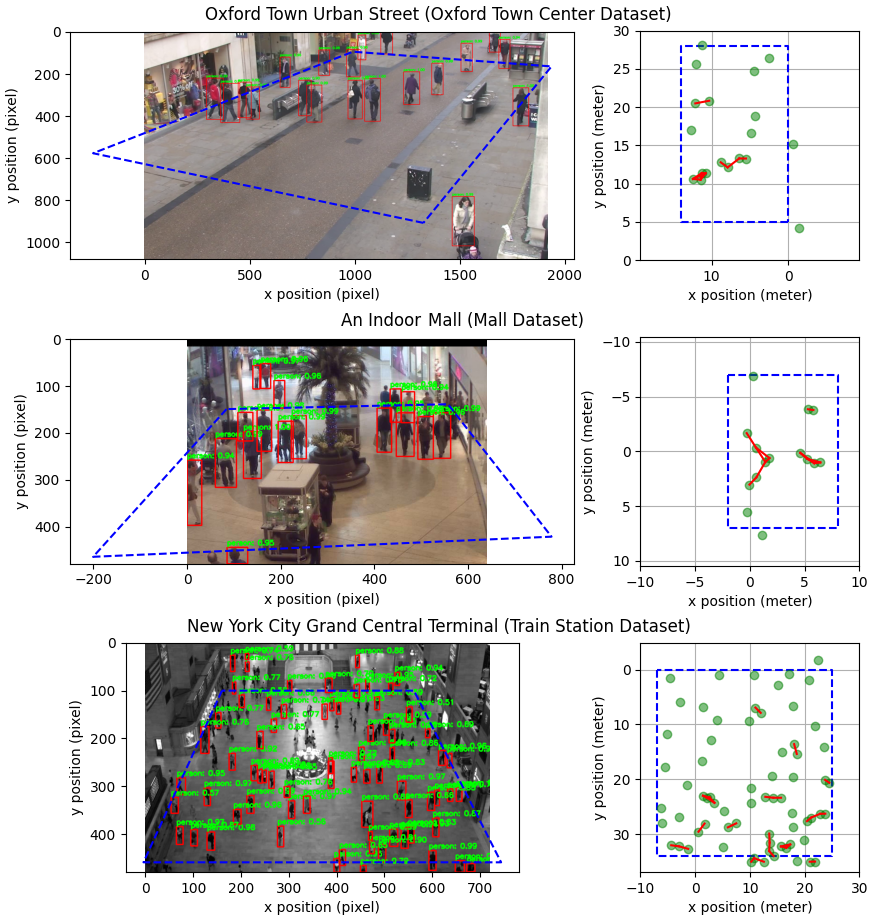}
	\caption{Illustration of pedestrian detection using Faster R-CNN~\cite{ren2015faster} and the corresponding social distancing.}
	\label{fig:detection}
\end{figure*}

\section{Experiments}

We conducted 3 case studies to evaluate the proposed method. Each case utilizes a different pedestrian crowd dataset. They are Oxford Town Center Dataset (an urban street)~\cite{benfold2011stable}, Mall Dataset (an indoor mall)~\cite{chen2012feature}, and Train Station Dataset (New York City Grand Central Terminal)~\cite{zhou2012understanding}. Table~\ref{tab:datasets} shows detailed information about these datasets.


\subsection{Implementation details}\label{sec:implementation}

The first step was finding the perspective transformation matrix $\mathbf{M}$ for each dataset. For Oxford Town Center Dataset, we directly used the transformation matrix available on its official website. For Train Station Dataset, we found the floor plan of NYC Grand Central Terminal and measured the exact distances among some key points that were used for calculating the perspective transformation. For Mall Dataset, we first estimated the size of a reference object in the image by comparing it with the width of detected pedestrians and then utilized the key points of the reference object to calculate the perspective transformation.

The second step was applying the pedestrian detector on each dataset. The experiments were conducted on a regular PC with an Intel Core i7-4790 CPU, 32GB RAM, and an Nvidia GeForce GTX 1070Ti GPU running Ubuntu 16.04 LTS 64-bit operating system. Once the pedestrians were detected, their positions were converted from the image coordinates into the real-world coordinates. 

The last step was conducting social distancing monitoring and finding the critical density $\rho_c$. Only the pedestrians within the ROI were considered. The statistics of the social density $\rho$, the inter-pedestrian distances $d_{i,j}$, and the number of violations $v$ were recorded over time. The analysis of statistics is described in the following section.



\begin{figure*}
	\includegraphics[width=1\linewidth]{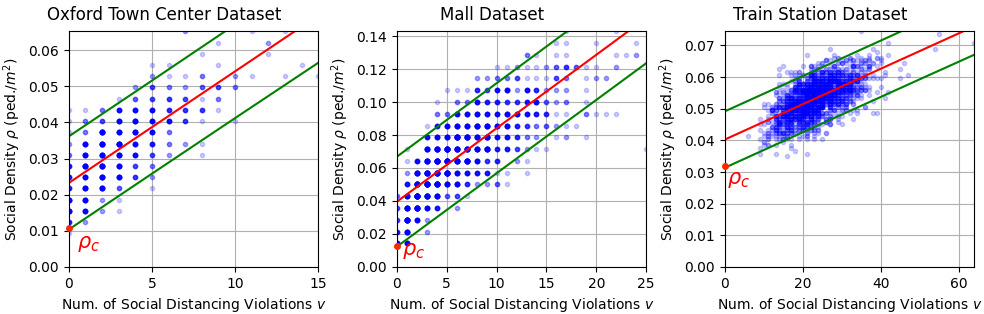}
	\caption{Linear regression (red line) of the social density $\rho$ versus number of social distancing violations $v$ data. Green lines indicate the prediction intervals. The critical social densities $\rho_c$ are the x-intercepts of the regression lines. }
	\label{fig:regression_density_vs_violation}
\end{figure*}

\section{Results}

\subsection{Pedestrian Detection}

We experimented with two different deep CNN based object detectors: Faster R-CNN and YOLOv4. Figure~\ref{fig:detection} shows the pedestrian detection results using Faster R-CNN~\cite{ren2015faster} and the corresponding social distancing in world coordinates. The detector performances are given in Table~\ref{tab:detectors}. As can be seen in the table, both detectors achieved real-time performance. In terms of detection accuracy, we provide the results of MS COCO dataset from the original works~\cite{ren2015faster, bochkovskiy2020yolov4}. 


\subsection{Social Distancing Monitoring}

\begin{figure*}
	\includegraphics[width=1\linewidth]{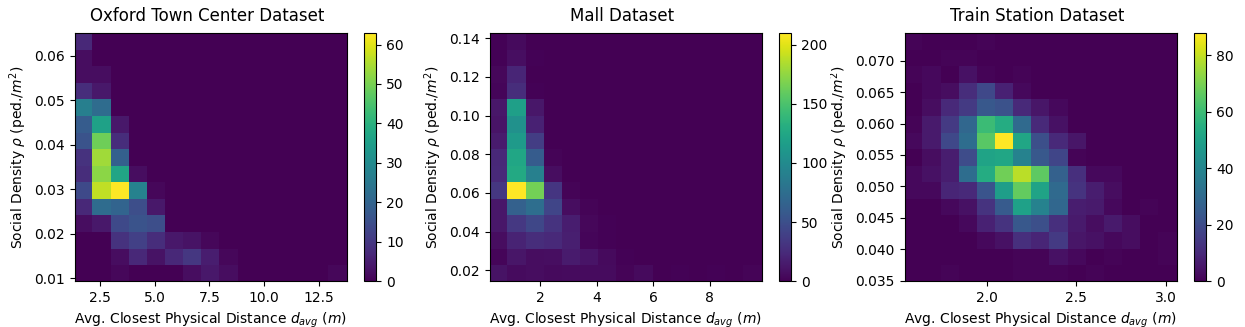}
	\caption{2D histograms of the social density $\rho$ versus the average closest physical distance $d_\text{avg}$. }
	\label{fig:hist_density_vs_avg_dists}
\end{figure*}

\begin{figure*}
    \centering
	\includegraphics[width=0.8\linewidth]{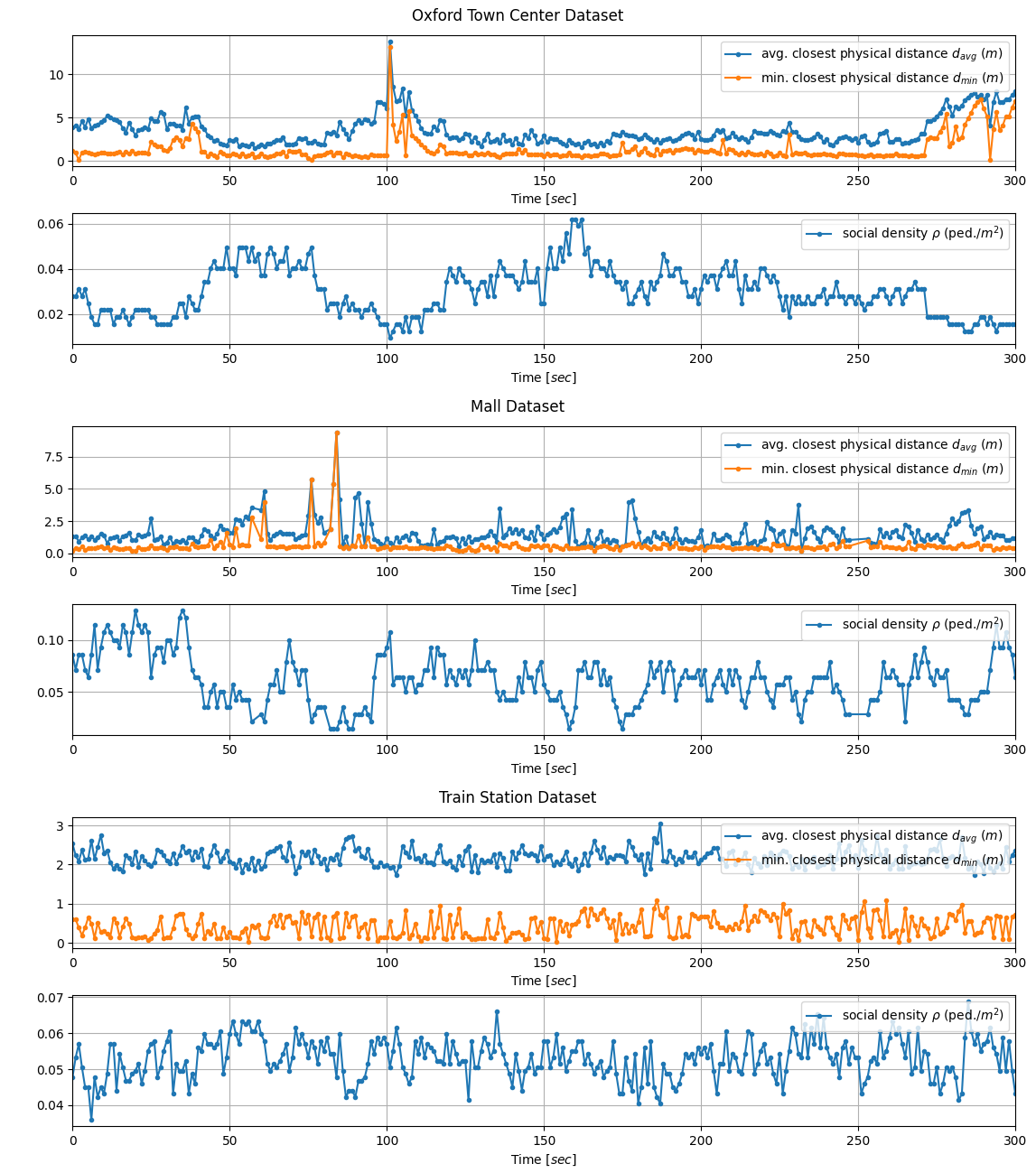}
	\caption{The change of minimum closest physical distance $d_\text{min}$, average closest physical distance $d_\text{avg}$, and the social density $\rho$ over time.}
	\label{fig:statistics_vs_time}
\end{figure*}

For pedestrian $i$, the closest physical distance is $d_i^\text{min}=\min(d_{i,j})$, $ \forall j \neq i \in \{1, 2, \cdots n\}$. Based on $d_i^\text{min}$, we further calculated two metrics for social distancing monitoring: the minimum closest physical distance $d_\text{min}=\min(d_i^\text{min}), \forall i \in \{1, 2, \cdots. n\}$ and the average closest physical distance $d_\text{avg} =\frac{1}{n}\sum_{i=1}^nd_i^\text{min}$.
Figure~\ref{fig:statistics_vs_time} shows the change of $d_\text{min}$ and $d_\text{avg}$ as time evolves. They are compared with the social density $\rho$. From the figure we can see that when $\rho$ is relatively low, both the $d_\text{min}$ and $d_\text{avg}$ are relatively high, for example, $t=85s$ in Train Station Dataset, $t=80s$ in Mall Dataset, and $t=100s$ in Oxford Town Center Dataset. This shows a clear negative correlation between $\rho$ and $d_\text{avg}$. The negative correlation is further visualized as 2D histograms in Figure~\ref{fig:hist_density_vs_avg_dists}.

\subsection{Critical Social Density}

\begin{figure*}
	\includegraphics[width=1\linewidth]{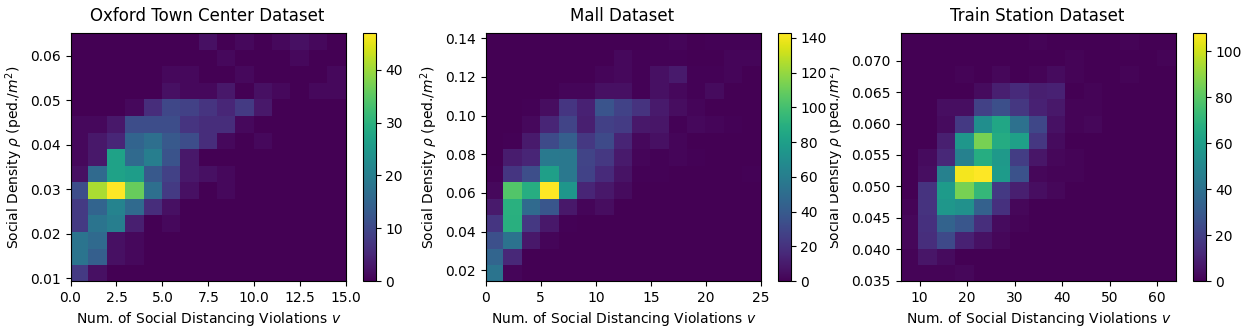}
	\caption{2D histograms of the social density $\rho$ versus the number of social distancing violations $v$. From the histograms we can see a linear relationship with positive correlation. }
	\label{fig:hist_density_vs_violation}
\end{figure*}

To find the critical density $\rho_c$, we first investigated the relationship between the number of social distancing violations $v$ and the social density $\rho$ in 2D histograms, as shown in Figure~\ref{fig:hist_density_vs_violation}. As can be seen in the figure, $v$ increases with an increase in $\rho$, which indicates a linear relationship with a positive correlation. 

Then, we conducted the simple linear regression, using the regression model of equation~(\ref{eq:regression}), on the data points of $v$ versus $\rho$. The skewness values of $\rho$ for Oxford Town Center Dataset, Mall Dataset, and Train Station Dataset are 0.25, 0.16, and -0.07, respectively, indicating the distributions of $\rho$ are symmetric. This satisfies the normality assumption of the error term in linear regression. The regression result is displayed in figure~\ref{fig:regression_density_vs_violation}. The critical density $\rho_c$ was identified as the lower bound of the prediction interval at $v=0$.

Table~\ref{tab:critical_density} summaries the identified critical densities $\rho_c$ as well as the intercepts $\beta_0$ of the regression models. The obtained critical density values for all datasets are similar. They also follow the patterns of the data points as illustrated in Figure~\ref{fig:regression_density_vs_violation}. This verified the effectiveness of our method. 
\begin{table}
    \begin{center}
        \begin{tabular}{|l|c|c|c|}
            \hline
            Dataset & FPS & Resolution & Duration\\
            \hline\hline
            Oxford Town Ctr. & 25       & $1920\times 1080$ & 5 mins\\
            Mall             & $\sim 1$ & $640\times 480$   & 33 mins\\
            Train Station    & 25       & $720\times 480$   & 33 mins\\
            \hline
        \end{tabular}
    \end{center}
    \caption{Information of each pedestrian dataset. }
    \label{tab:datasets}
\end{table}

\begin{table}
    \begin{center}
        \begin{tabular}{|l|c|c|}
            \hline
            Method       & mAP (\%) & Inference Time (sec)\\
            \hline\hline
            Faster R-CNN~\cite{ren2015faster}   & 42.1-42.7 & 0.145 / 0.116 / 0.108\\
            YOLOv4~\cite{bochkovskiy2020yolov4} & 41.2-43.5 & 0.048 / 0.050 / 0.050 \\
            \hline
        \end{tabular}
    \end{center}
    \caption{The real-time performance of pedestrian detectors. The inference time reports the mean inference time for Oxford Town Center / Train Station / Mall datasets, respectively.}
    \label{tab:detectors}
\end{table}

\begin{table}
    \begin{center}
        \begin{tabular}{|l|c|c|c|}
            \hline
            Dataset & Intercept $\beta_0$ & Critical Density $\rho_c$ \\
            \hline\hline
            Oxford Town Ctr. & 0.0233  & 0.0104 \\
            Mall             & 0.0396  & 0.0123 \\
            Train Station    & 0.0403  & 0.0314 \\
            \hline
        \end{tabular}
    \end{center}
    \caption{Critical social density of each dataset. The critical density was identified as the lower bound of the prediction interval at the number of social distancing violations $v=0$.}
    \label{tab:critical_density}
\end{table}



\section{Conclusion}

This work proposed an AI and monocular camera based real-time system to monitor the social distancing. We use the proposed critical social density to avoid overcrowding by modulating inflow to the ROI. The proposed method was verified using 3 different pedestrian crowd datasets.

There are some missing detections in the Train Station Dataset, as in some areas the pedestrian density is extremely high and occlusion happens. However, after some qualitative analysis, we concluded that most of the pedestrians were captured and the idea of finding critical social density is still valid.

Pedestrians who belong to a group were not considered as a group in the current work, which can be a future direction. Nevertheless, one may argue that even individuals who have close relationships should still try to practice social distancing in public areas.


{\small
	\bibliographystyle{IEEEtran}

	\bibliography{egbib}
}
	
\end{document}